\documentclass[graybox]{svmult}

\usepackage{amsmath,amscd}%
\usepackage{amsfonts}%
\usepackage{amssymb}%
\usepackage[heads=vee]{diagrams}

\newcommand{\reconstruction}{\mathcal{I}} 
\newcommand{\reduction}{\mathcal{R}} 
\newcommand{\projection}{\pi_{h}} 

\newcommand{\ud}{\mathrm{d}} 
\newcommand{\p}{\partial}
\newcommand{\grad}{\mathrm{grad}\,} 
\newcommand{\curl}{\mathrm{curl}\,} 
\renewcommand{\div}{\mathrm{div}\,} 

\newcommand{\figref}[1]{Figure~\ref{#1}} 

\newcommand{\norm}[1]{\Vert #1\Vert}

\usepackage{mathptmx}       
\usepackage{helvet}         
\usepackage{courier}        
\usepackage{type1cm}        
\usepackage{makeidx}         
\usepackage{graphicx}        
\usepackage{multicol}        
\usepackage[bottom]{footmisc}

\makeindex             

\begin{document}

\title*{Higher-order compatible discretization on hexahedrals}
\author{Jasper Kreeft and Marc Gerritsma}
\institute{Jasper Kreeft \at Shell Global Solutions, Grasweg 31, 1031 HW, Amsterdam, \email{jasper.kreeft@shell.com}
\and Marc Gerritsma \at Delft University of Technology, Kluyverweg 2, 2629 HT, Delft, \email{m.i.gerritsma@tudelft.nl}}
%
%
\maketitle

\abstract*{We derive a compatible discretization method that relies heavily on the underlying geometric structure and obeys the topological sequences and commuting properties that are constructed. As a sample problem we consider the vorticity-velocity-pressure formulation of the Stokes problem. We motivate the choice for a mixed variational formulation based on both geometric as well as physical arguments. Numerical tests confirm the theoretical results that we obtain a pointwise divergence-free solution for the Stokes problem and that the method obtains optimal convergence rates.}

\abstract{We derive a compatible discretization method that relies heavily on the underlying geometric structure, and obeys the topological sequences and commuting properties that are constructed. As a sample problem we consider the vorticity-velocity-pressure formulation of the Stokes problem. We motivate the choice for a mixed variational formulation based on both geometric as well as physical arguments. Numerical tests confirm the theoretical results that we obtain a pointwise divergence-free solution for the Stokes problem and that the method obtains optimal convergence rates.}

\section{Introduction}\label{sec:introduction}
As sample problem we consider the Stokes flow problem in its vorticity-velocity-pressure formulation,
\begin{subequations}
\label{stokessinglevector}
\begin{align}
\vec{\omega}-\mathrm{curl}\,\vec{u}&=0\quad\mathrm{in}\ \Omega,\label{stokessinglevector1}\\
\mathrm{curl}\,\vec{\omega}+\mathrm{grad}\,p&=\vec{f}\quad\ \mathrm{in}\ \Omega,\label{stokessinglevector2}\\
\mathrm{div}\,\vec{u}&=0\quad\mathrm{in}\ \Omega.\label{stokessinglevector3}
\end{align}
\end{subequations}
In this article we consider prescribed velocity boundary conditions, $\vec{u}=0$ on $\p\Omega$, but the method holds for all admissible types of boundary conditions, see \cite{kreefterrorestimate}.

Despite the simple appearance of Stokes flow model, there exists a large number of numerical methods to simulate Stokes flow. They all reduce to two classes, that is, either circumventing the LBB stability condition, like stabilized methods, e.g. \cite{hughes1986}, or satisfying this condition, as in compatible or mixed methods, e.g. \cite{brezzifortin}. The last requires the construction of dedicated discrete vector spaces. Best known are the curl conforming N\'ed\'elec and divergence conforming Raviart-Thomas spaces. Here, we consider a subclass of compatible methods, i.e. {\em mimetic methods}. Mimetic methods do not solely search for appropriate vector spaces, but aim to mimic structures and symmetries of the continuous problem, see \cite{bochevhyman2006,brezzibuffa2010,kreeftpalhagerritsma2011,perot2011}. As a consequence of this mimicking, mimetic methods automatically preserve most of the physical and mathematical structures of the continuous formulation, among others the LBB condition and, most important, a pointwise divergence-free solution, \cite{kreefterrorestimate,kreeft2012}.

\begin{figure}[htbp]
\centering
\includegraphics[width=0.9\textwidth]{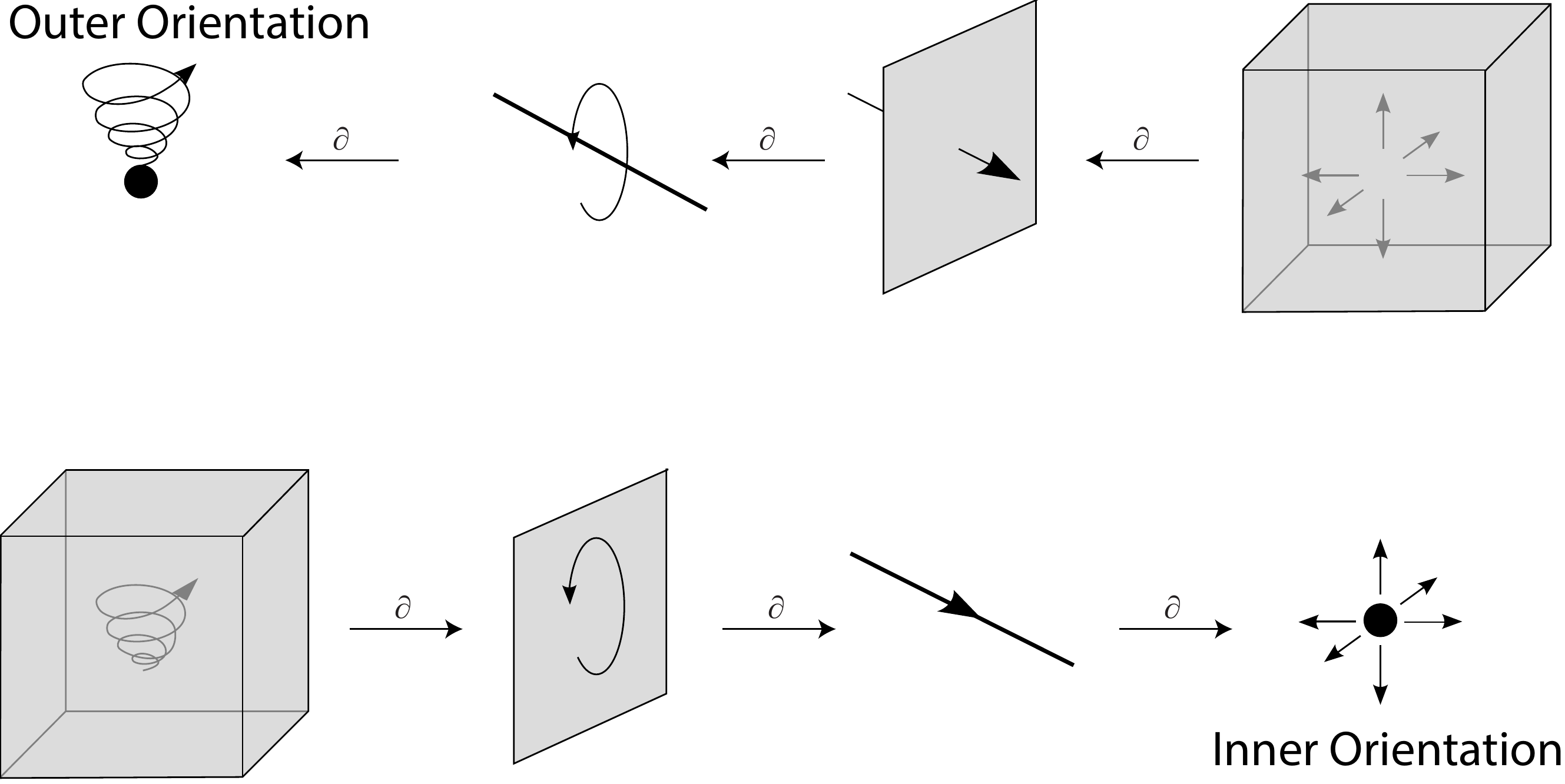}
\caption{The four geometric objects possible in $\mathbb{R}^3$, point, line, surface and volume, with outer- (above) and inner- (below) orientation. The boundary operator, $\partial$, maps $k$-dimensional objects to $(k-1)$-dimensional objects.}
\label{fig:manifoldswithorientation}
\end{figure}

At the heart of the mimetic method there are the well-known integral theorems of Newton-Leibniz, Stokes and Gauss, which couple the operators grad, curl and div, to the action of the boundary operator on a manifold. Therefore, obeying geometry and orientation will result in satisfying exactly the mentioned theorems, and consequently performing the vector operators exactly in a finite dimensional setting. In 3D we distinguish between four types of sub-manifolds, that is, points, lines, surfaces and volumes, and two types of orientation, namely, outer- and inner-orientation. Examples of sub-manifolds are shown in \figref{fig:manifoldswithorientation} together with the action of the boundary operator.

By creating a quadrilateral or hexahedral mesh, we divide the physical domain in a large number of these geometric objects, and to each geometric object we associate a discrete unknown. This implies that these discrete unknowns are \emph{integral quantities}. Since the three earlier mentioned theorems are integral equations, it follows for example that taking a divergence in a volume is equivalent to taking the sum of the integral quantities associated to the surrounding surface elements, i.e. the fluxes. So using integral quantities as degrees of freedom to perform a grad, curl or div, is equivalent to taking the sum of the degrees of freedom located at its boundary.

These relations are of purely topological nature. They form a topological sequence or complex. This sequence is fundamental. It has a direct connection with the complexes that are related to the physical domain, the computational domain, the physical problem and the discretization.

Although the original work, \cite{kreeft2012,kreeftpalhagerritsma2011}, was presented in terms of differential geometry and algebraic topology, here we will use vector calculus because it is the more common mathematical language. Nevertheless, we will put emphasis on the distinction between topology and metric, on complexes and on commuting diagrams, which drives the former two languages.

We make use of spectral element interpolation functions as basis functions. In the past nodal spectral elements were mostly used in combination with Galerkin projection (GSEM). The GSEM satisfies the LBB condition by lowering the polynomial degree of the pressure by two with respect to the velocity. This results in a method that is only weakly divergence-free, meaning that the divergence of the velocity field only convergence to zero with  mesh refinement. The present study uses mimetic spectral element interpolation or basis functions, \cite{kreeftpalhagerritsma2011}. The mixed mimetic spectral element method (MMSEM) satisfies the LBB condition and gives a pointwise divergence-free solution for all mesh sizes.

\section{Can we really discretize exactly?}\label{sec:stokes}
Since the Stokes flow model \eqref{stokessinglevector} should hold on a certain physical domain, we will include geometry by means of integration. In that case we can relate every physical quantity to a geometric object. Starting with the incompressibility constraint \eqref{stokessinglevector3} we have due to Gauss' divergence theorem,
\[
\int_V\div\vec{u}\,\ud V=\int_{\p V}\vec{u}\cdot\vec{n}\,\ud S=0,
\]
and using Stokes' circulation theorem the relation \eqref{stokessinglevector1} can be written as
\[
\int_S\vec{\omega}\times\vec{n}\,\ud S=\int_S\curl\vec{u}\times\vec{n}\,\ud S=\int_{\p S}\vec{u}\cdot\vec{t}\,\ud l.
\]
From the first relation it follows that $\div\vec{u}$ is associated to volumes. The association to a geometric object for velocity $\vec{u}$ is less clear. In fact it can be associated to two different types of geometric objects. 
A representation of velocity compatible with the incompressibility constraint is given in terms of the velocity flux, $\vec{u}\cdot\vec{n}$, \emph{through a surface} that bounds the volume, while in the circulation relation velocity, $\vec{u}\cdot\vec{t}$, is represented \emph{along a line} that bounds the surface. We will call the velocity vector through a surface {\em outer-oriented} and the velocity along a line segment {\em inner-oriented}. A similar distinction can be made for vorticity, see \cite{kreeft2012}.

\begin{figure}[htbp]
\centering
\includegraphics[width=0.9\textwidth]{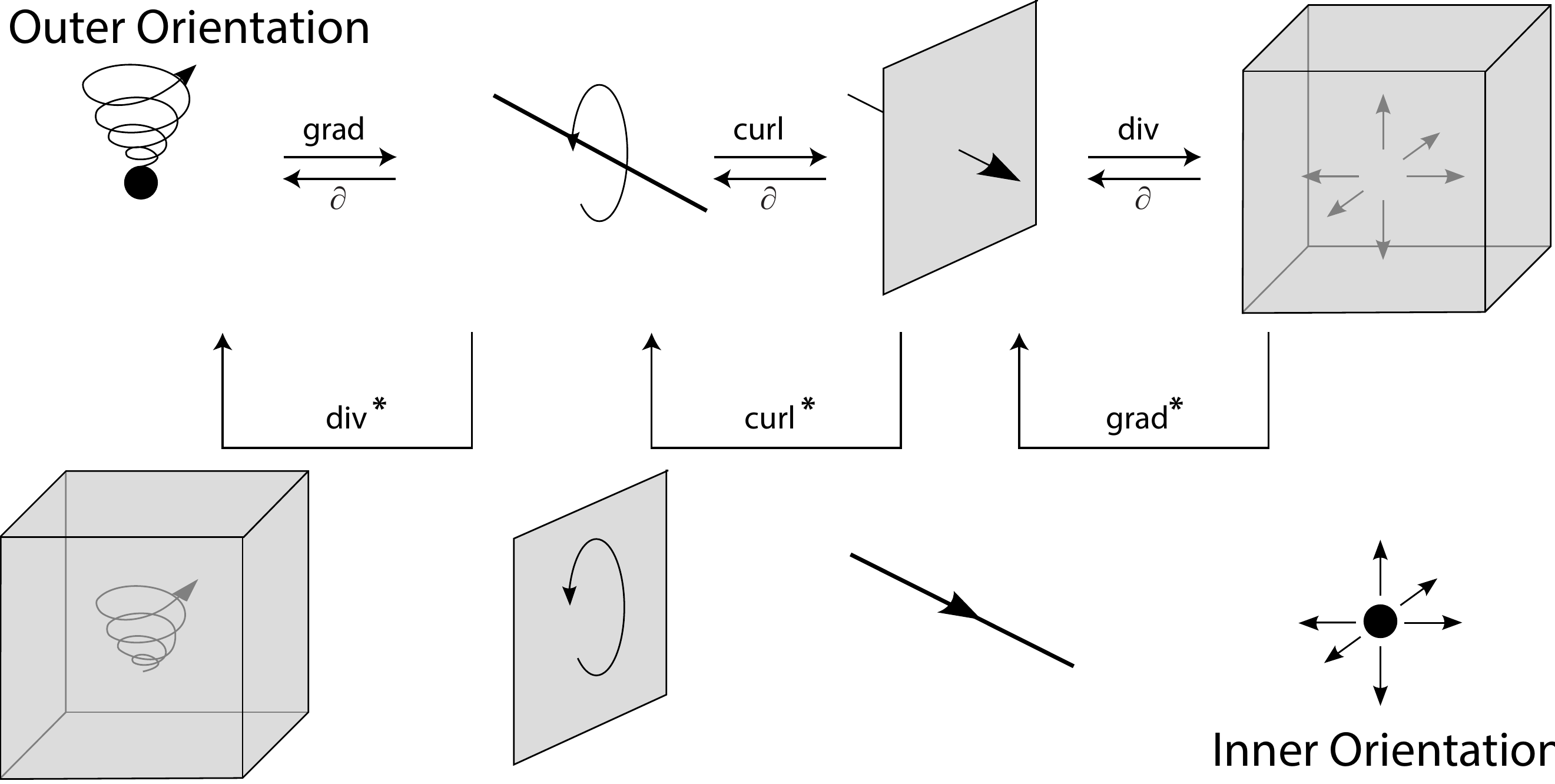}
\caption{Geometric interpretation of the action of the boundary operators, vector differential operators and their formal Hilbert adjoint operators.}
\label{fig:manifoldswithorientation2}
\end{figure}

The last equation to be considered is \eqref{stokessinglevector2}. This equation shows that classical Newton-Leibniz, Stokes circulation and Gauss divergence theorems tell only half the story. From the perspective of the classical Newton-Leibniz theorem, the gradient acting on the pressure relates line values to their corresponding end point, while the Stokes circulation theorem shows that the curl acting on the vorticity vector relates surface values to the line segment enclosing it. So how does this fit into one equation? In fact, from a geometric perspective, there exists two gradients, two curls and two divergence operators. One of each is related to the mentioned integral theorems as explained above. The others are their formal adjoint operators. Let grad, curl and div be the original differential operators associated to the mentioned integral theorems, then the formal Hilbert adjoint operators grad$^*$, curl$^*$ and div$^*$ are defined as,
\[
\big(\vec{a},-\mathrm{grad}^*\,b\big)_\Omega:=\big(\div\vec{a},b\big)_\Omega,\ \big(\vec{a},\mathrm{curl}^*\,\vec{b}\big)_\Omega:=\big(\curl\vec{a},\vec{b}\big)_\Omega,\  \big(a,-\mathrm{div}^*\,\vec{b}\big)_\Omega:=\big(\grad a,\vec{b}\big)_\Omega.
\]
From a geometric interpretation, the adjoint operators detours via the opposite type of orientation. Where div relates a vector quantity associated to surfaces to a scalar quantity associated to a volume enclosed by these surfaces. Its adjoint operator, grad$^*$, relates a scalar quantity associated with a volume to a vector quantity associated with its surrounding surfaces. This is illustrated in \figref{fig:manifoldswithorientation2}. Following \figref{fig:manifoldswithorientation2}, the adjoint operator grad$^*$ consists of three consecutive steps: First, switch from an outer oriented scalar associated to volumes to an  inner oriented scalar associated to points, then take the derivative and finally switch from an inner oriented vector associated to lines to an outer oriented vector associated to surfaces. In a similar way we can describe the derivatives curl$^*$ and div$^*$. 

Since the horizontal relations are purely topological and the vertical relations purely metric, the operators grad, curl and div are purely topological operators, while grad$^*$, curl$^*$ and div$^*$ are metric. This makes them much harder to discretize.

Now \eqref{stokessinglevector2} could then either be associated to an inner-oriented line segment by rewriting it as
\[
\mathrm{curl}^*\,\vec{\omega}+\grad p=\vec{f},
\]
or be associated to an outer-oriented surface by rewriting it as
\[
\curl\vec{\omega}+\mathrm{grad}^* p=\vec{f}.
\]
Without geometric considerations we could never make a distinction between grad, curl and div and their associated Hilbert adjoints div$^*$, curl$^*$ and grad$^*$.

Since our focus is on obtaining a pointwise divergence-free discretization, we decide to use the expression where the equations are associated to outer-oriented geometric objects,
\begin{subequations}
\begin{align}
\vec{\omega}-\mathrm{curl}^*\,\vec{u}&=0\quad\mathrm{in}\ \Omega,\\
\curl\vec{\omega}+\mathrm{grad}^*\,p&=\vec{f}\quad\ \mathrm{in}\ \Omega,\\
\div\vec{u}&=0\quad\mathrm{in}\ \Omega,
\end{align}
\label{eq:system_vector_calculus}
\end{subequations}
where the first equation is associated to outer-oriented line segments, the second to outer-oriented surfaces and the third to outer-oriented volumes.

\section{Complexes}
\figref{fig:manifoldswithorientation2} reveals already a number of sequence or complex structures. Starting from geometry, we consider points, $P$, lines, $L$, surfaces, $S$, and volumes, $V$. They possess a sequence in combination with the boundary operator, $\p$. The boundary of a volume is a surface, the boundary of a surface is a line and the boundary of a line are its two end points. This results in the following complex,
\begin{equation}
0\stackrel{\p}{\longleftarrow}P\stackrel{\p}{\longleftarrow}L\stackrel{\p}{\longleftarrow}S\stackrel{\p}{\longleftarrow}V.
\end{equation}
An important property of the complex is that if we apply the boundary operator twice, we always find an empty set, e.g. if $S=\p V$, then $\p S=\emptyset$. As follows directly from the previously mentioned integral theorems, it follows, as a consequence of $\p\p=\emptyset$, that $\curl\grad=0$ and $\div\curl=0$. The derivatives themselves also form a complex. In a Hilbert setting this becomes,
\begin{equation}
H^1(\Omega)\stackrel{\grad}{\longrightarrow}H(\Omega,\mathrm{curl})\stackrel{\curl}{\longrightarrow}H(\Omega,\mathrm{div})\stackrel{\div}{\longrightarrow}L^2(\Omega),
\end{equation}
and using the Hilbert adjoint relations we also obtain the adjoint complex with properties $\mathrm{curl}^*\mathrm{grad}^*=0$ and $\mathrm{div}^*\mathrm{curl}^*=0$,
\begin{equation}
L^2_0(\Omega)\stackrel{-\mathrm{div}^*}{\longleftarrow}H_0(\Omega,\mathrm{div}^*)\stackrel{\mathrm{curl}^*}{\longleftarrow}H_0(\Omega,\mathrm{curl}^*)\stackrel{-\mathrm{grad}^*}{\longleftarrow}H^1_0(\Omega)
\end{equation}
In the Hilbert setting, the variables of the Stokes problem are in the following spaces, $\vec{\omega}\in H(\Omega,\mathrm{curl})\cap H_0(\Omega,\mathrm{div}^*)$, $\vec{u}\in H(\Omega,\mathrm{div})\cap H_0(\Omega,\mathrm{curl}^*)$ and $p\in L^2(\Omega)\cap H_0(\Omega,\mathrm{grad}^*)$.
It is hard, if even possible at all, to find discrete vector spaces that are subsets of these function spaces and simultaneously satisfy the complex properties. Instead, the Stokes problem can be cast into an equivalent variational or mixed formulation where we make use of the Hilbert adjoint properties. This simplifies the function spaces of the flow variables. The mixed formulation reads; 
\begin{svgraybox}
Find $(\vec{\omega},\vec{u},p)\in\{H(\Omega,\mathrm{curl})\times H(\Omega,\mathrm{div})\times L^2(\Omega)\}$ with $\vec{f}\in H(\Omega,\mathrm{div})$ given, for all $(\vec{\sigma},\vec{v},q)\in\{H(\Omega,\mathrm{curl})\times H(\Omega,\mathrm{div})\times L^2(\Omega)\}$, such that,
\begin{subequations}
\begin{align}
\big(\vec{\sigma},\vec{\omega}\big)_\Omega-\big(\curl\vec{\sigma},\vec{u}\big)_\Omega&=0,\\
\big(\vec{v},\curl\vec{\omega}\big)_\Omega-\big(\div\vec{v},p\big)_\Omega&=\big(\vec{v},\vec{f}\big)_\Omega,\\
\big(q,\div\vec{u}\big)_\Omega&=0.
\end{align}
\end{subequations}
\end{svgraybox}
With the formulation and corresponding function spaces, we are able to construct compatible discrete vector spaces. Note that we now completely avoid the metric dependent derivatives grad$^*$ and curl$^*$, and their corresponding complex.

\vspace{-0.5cm}
\section{Discretization of Stokes problem}\label{sec:discretization}

\textbf{Degrees of freedom.} In many numerical methods, especially in finite difference and finite element methods, the discrete coefficients are point values. In the proposed mimetic structure, the discrete unknowns represent integral values on $k$-dimensional submanifolds, ranging from points to volumes, so $0\leq k\leq 3$. These $k$-dimensional submanifolds are oriented, constitute the computational domain and span the physical domain.
The concept of orientation shown in Figures \ref{fig:manifoldswithorientation} and \ref{fig:manifoldswithorientation2} gave rise to the boundary operator, $\p$, which can be represented by connectivities consisting only of -1, 0 and 1, see also \cite{kreeft2012}.

\begin{figure}[htb]
\centering
\includegraphics[width=0.9\textwidth]{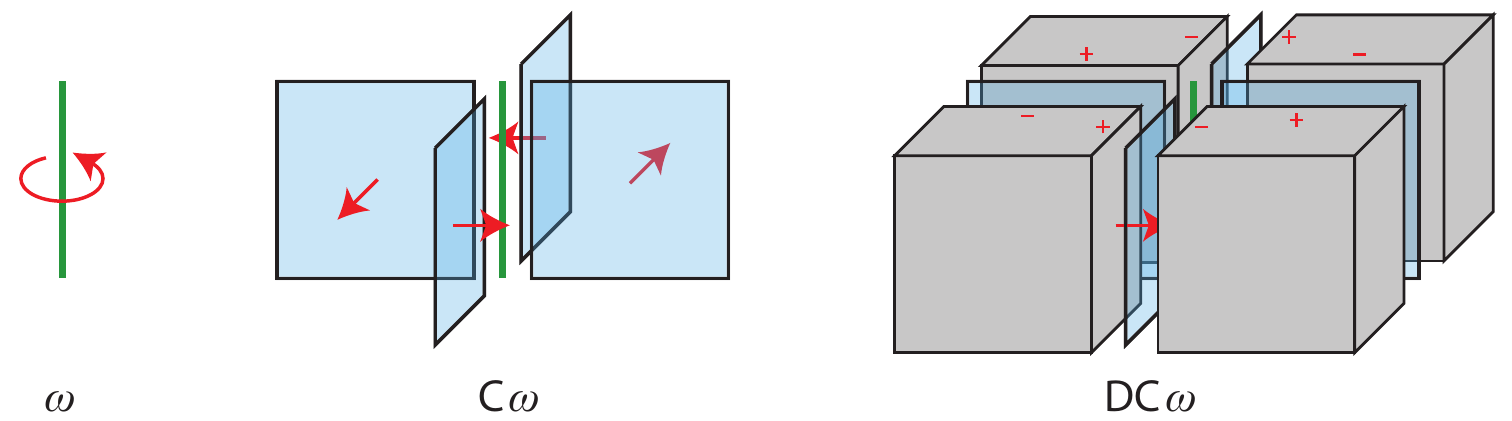}
\caption{The action of twice the coboundary operator $\delta$ on a vorticity d.o.f. has a zero net result on its surrounding volumes, because they all have both a positive and a negative contribution from its neighboring velocity faces.}
\label{fig:coboundarycoboundary}
\end{figure}

The space of degrees of freedom are given by $\mathcal{P}$, $\mathcal{L}$, $\mathcal{S}$ and $\mathcal{V}$. These spaces form a duality pairing with the geometric spaces $P$, $L$, $S$ and $V$. The degrees of freedom are integral values, i.e.
\begin{equation}
\int_l\vec{w}\cdot\vec{t}\,\ud l\ \in\mathcal{L},\quad \int_S \vec{u}\cdot\vec{n}\,\ud S\ \in\mathcal{S},\quad \int_V p\,\ud V\ \in\mathcal{V}.
\end{equation}
By the definition of the degrees of freedom spaces and the previously mentioned integral theorems, we can define the formal adjoint of the boundary operator, i.e. the coboundary operator, $\delta$. The coboundary operator is the discrete representation of the topological derivatives grad, curl and div. Since $\p\p=\emptyset$, it follows from a discrete Newton-Leibniz, Stokes and Gauss theorem that applying the coboundary operator twice  is always zero, $\delta\delta=\emptyset$ (see \cite{bochevhyman2006,kreeft2012}). The coboundary operator also has matrix representations, $\mathsf{G}$, $\mathsf{C}$ and $\mathsf{D}$, that are the transpose of the connectivity matrices. We obtain the following topological sequence,
\begin{equation}
\mathcal{P}\stackrel{\mathsf{G}}{\longrightarrow}\mathcal{L}\stackrel{\mathsf{C}}{\longrightarrow}\mathcal{S}\stackrel{\mathsf{D}}{\longrightarrow}\mathcal{V},
\end{equation}
where $\mathsf{CG}=\vec{0}$ and $\mathsf{DC}=\vec{0}$. These matrices will explicitly appear in the final matrix system. An illustration of $\mathsf{DC}=\vec{0}$ is given in \figref{fig:coboundarycoboundary}.
More details on the structure of geometry, orientation and degrees of freedom can be found in Gerritsma et al, \cite{gerritsmaetal2012}.

\textbf{Mimetic Operators.} Let $W=H(\Omega,\mathrm{curl})$, $V=H(\Omega,\mathrm{div})$ and $Q=L^2(\Omega)$. The discretization of the flow variables involves a projection operator, $\pi_h$, from the complete vector spaces $W$, $V$ and $Q$, to the discrete vector spaces $W_h$, $V_h$ and $Q_h$. Here the flow variables are expressed in terms of d.o.f. defined on $k$-cells, and corresponding interpolation functions (also called basis-functions). The projection operator actually consists of two steps, a reduction operator, $\reduction$, that integrates the flow variables on $k$-cells, and a reconstruction operator, $\reconstruction$, that interpolates the d.o.f. using the appropriate basis-functions. These mimetic operators were defined in \cite{bochevhyman2006,kreeftpalhagerritsma2011}. A composition of the two operators gives the projection operator $\projection=\reconstruction\circ\reduction$\footnote{For completeness, in a Hilbert setting the projection needs an additional smoothing argument. This step is ignored here to increase readibility. See \cite{kreefterrorestimate} for more details.}.

Reduction operator $\mathcal{R}$ is simply defined by integration. It possesses the following commutation relations,
\begin{equation}
\mathcal{R}\grad=\mathsf{G}\mathcal{R},\quad\mathcal{R}\curl=\mathsf{C}\mathcal{R},\quad\mathcal{R}\div=\mathsf{D}\mathcal{R}.
\end{equation}
The treatment of the reconstruction operator leaves some freedom, as long as it satisfies the following properties: be the right inverse of the reduction, $\mathcal{RI}=Id$, be the approximate left inverse of the reduction, $\mathcal{IR}=Id+\mathcal{O}(h^p)$, and it should possess the following commutation relations,
\begin{equation}
\grad\mathcal{I}=\mathcal{I}\mathsf{G},\quad\curl\mathcal{I}=\mathcal{I}\mathsf{C},\quad\div\mathcal{I}=\mathcal{I}\mathsf{D}.
\end{equation}
\begin{svgraybox}
When both the reduction and reconstruction operators commute with continuous and discrete differentiation, than also the projection operator $\pi_h$ possesses a commutation relation with differentiation. In case of the divergence operator, which is relevant to obtain a pointwise divergence-free solution, the commutation relation is given by,
\begin{equation}
\div\pi_h=\div\mathcal{IR}=\mathcal{I}\mathsf{D}\mathcal{R}=\mathcal{IR}\div=\pi_h\div.
\label{commutationprojection}
\end{equation}
The commutation relations in case of divergence are illustrated below,
\[
\begin{CD}
V @>\div>> Q\\
@VV\reduction V @VV\reduction V\\
\mathcal{S} @>\mathsf{D}>> \mathcal{V}.
\end{CD}
\quad+\quad
\begin{CD}
\mathcal{S} @>\mathsf{D}>> \mathcal{V}\\
@VV\reconstruction V @VV\reconstruction V\\
V_h @>\div>> Q_h.
\end{CD}
\quad=\quad
\begin{CD}
V @>\div>> Q\\
@VV\projection V @VV\projection V\\
V_h @>\div>> Q_h.
\end{CD}
\]
Since property \eqref{commutationprojection} also holds for the grad and curl, we obtain the following complex for discrete vector spaces,
\begin{equation}
\Phi_h\stackrel{\grad}{\longrightarrow}W_h\stackrel{\curl}{\longrightarrow}V_h\stackrel{\div}{\longrightarrow}Q_h.
\end{equation}
\end{svgraybox}
In practice we use $\mathcal{I}\mathsf{D}\mathcal{R}$ from \eqref{commutationprojection} in computations. Relation \eqref{commutationprojection} implies among others that it satisfies the discrete LBB condition,
\begin{equation}
\beta_h:=\inf_{q_h\in Q_h}\sup_{v_h\in V_h}\frac{\big(q_h,\div\vec{v_h}\big)_\Omega}{\norm{q_h}_{Q}\norm{v_h}_{V}}>\beta>0,
\end{equation}
where $\beta$ is the inf-sup constant of the continuous problem, \eqref{stokessinglevector}. Whereas the LBB condition is a measurement for numerical stability, the commutation relation indicates physical correctness of the numerical method. This last is a much  stronger statement, which includes also the former.

The conditions on the reconstruction operator have led to the construction of mimetic spectral element basis-functions, \cite{gerritsma2011,kreeftpalhagerritsma2011}. Since we use a tensor-based construction of point, line, surface and volume corresponding basis-functions, we only need nodal and edge interpolation functions. The nodal interpolation functions are the well-known Lagrange polynomials. The edge polynomials were derived from the Lagrange polynomials, based on the given conditions. For a set of Lagrange polynomials, $l_i(x)$, $i=0,\hdots,N$, the edge polynomials, $e_i(x)$, $i=1,\hdots,N$, are given by,
\begin{equation}
e_i(x)=-\sum_{k=0}^{i-1}\frac{\ud l_k(x)}{\ud x}.
\end{equation}
The Lagrange and edge polynomials possess the condition $\mathcal{RI}=Id$, i.e.,
\begin{equation}
l_i(x_j)=\delta_{i,j},\quad\quad \int_{x_{j-1}}^{x_j}e_i(x)\,\ud x=\delta_{i,j},
\end{equation}
where $\delta_{i,j}$ is the Kronecker delta. The interpolation function for a variable associated to a surface, for example, is given by,
$s_{i,j,k}(x,y,z)=\left\{l_i(x)e_j(y)e_k(z),e_i(x)l_j(y)e_k(z),\right.$
$\left.e_i(x)e_j(y)l_k(z)\right\}$.

\begin{svgraybox}
\begin{example}[\textbf{Divergence operator in 2D}]\label{ex:div}
One of the most interesting properties of the mimetic method presented in this paper, is that within our weak formulation, the divergence-free constraint is satisfied pointwise. Let $u_h\in V_h$ be the velocity flux defined as
\begin{equation}
\vec{u}_h=
\begin{pmatrix}
\sum_{i=0}^N\sum_{j=1}^Nu_{i,j}l_i(x)e_j(y)\\
\sum_{i=1}^N\sum_{j=0}^Nv_{i,j}e_i(x)l_j(y)
\end{pmatrix}
.
\end{equation}
Then the change of mass, $m_h\in Q_h$, is equal to the divergence of $\vec{u}_h$,
\begin{align}
m_h=\div\vec{u}_h&=\sum_{i=1}^N\sum_{j=1}^N(u_{i,j}-u_{i-1,j}+v_{i,j}-v_{i,j-1})e_i(x)e_j(y).\nonumber\\
&=\sum_{i=1}^N\sum_{j=1}^Nm_{i,j}e_i(x)e_j(y),
\end{align}
where $m_{i,j}=u_{i,j}-u_{i-1,j}+v_{i,j}-v_{i,j-1}$ can be compactly written as $\mathsf{m}=\mathsf{Du}$. Note that if the mass production is zero, as in our model problem \eqref{stokessinglevector3}, the incompressibility constraint can already be satisfied at the discrete level. Interpolation using $e_i(x)e_j(y)$ then results in a solution of velocity $\vec{u}_h$ that is pointwise divergence-free.
\end{example}
\end{svgraybox}

\section{A priori error estimates}
By standard interpolation theory it follows that we obtain the following $h$-convergence rates for the interpolation errors of the flow variables,
\begin{equation}
\norm{\vec{\omega}-\pi_h\vec{\omega}}_{H(\mathrm{curl})}=\mathcal{O}(h^N),\quad \norm{\vec{u}-\pi_h\vec{u}}_{H(\mathrm{div})}=\mathcal{O}(h^N),\quad
\norm{p-\pi_hp}_{L^2}=\mathcal{O}(h^N),
\end{equation}
and that $\norm{\div\vec{u}-\div\pi_h\vec{u}}_{L^2}=0$ due to the commuting property.

In cases with empty harmonic vector spaces, we have that the discrete vector spaces are conforming, i.e., $W_h\subset W$, $V_h\subset V$ and $Q_h\subset Q$. Moreover, due to the commuting property, it follows that these spaces are compatible, i.e., $\curl W_h\subset V_h$ and $\div V_h=Q_h$. Finally they possess a Helmholtz-Hodge decomposition, $\vec{\sigma}=\grad\phi+\mathrm{curl}^*\vec{v}$ and $\vec{v}=\curl\vec{\sigma}+\mathrm{grad}^*q$. In terms of vector spaces, this is, $W_h=Z_{W_h}\oplus Z_{W_h}^\perp$ and $V_h=Z_{V_h}\oplus Z_{V_h}^\perp$, where $Z$ refers to the kernel or nullspace and $Z^\perp$ to its orthogonal complement. Having all these properties, a priori error estimates are derived in \cite{kreefterrorestimate} that show optimal convergence rates for all admissible boundary conditions, including the no-slip boundary condition, which is non-trivial in mixed finite element methods. The a priori error estimates are given by
\begin{gather}
\norm{\vec{\omega}-\vec{\omega}_h}_W\leq C\inf_{\vec{\sigma}_h\in W_h}\norm{\vec{\omega}-\vec{\sigma}_h}_W,\\
\norm{\vec{u}-\vec{u}_h}_V\leq C\inf_{\vec{v}_h\in V_h}\norm{\vec{u}-\vec{v}_h}_V + C\inf_{\vec{\sigma}_h\in W_h}\norm{\vec{\omega}-\vec{\sigma}_h}_W,\\
\norm{p-p_h}_Q\leq C\inf_{q_h\in Q_h}\norm{p-q_h}_Q + C\inf_{\vec{v}_h\in V_h}\norm{\vec{u}-\vec{v}_h}_V + C\inf_{\vec{\sigma}_h\in W_h}\norm{\vec{\omega}-\vec{\sigma}_h}_W,
\end{gather}
where the constants $C$ will differ in each case and are independent of $h$. It shows that the rate of convergence of the approximation errors are the same as those of the interpolation errors.


\section{Numerical Results}\label{sec:numericalresults}
For many years, the lid-driven cavity flow was considered one of the classical benchmark cases for the assessment of numerical methods and the verification of incompressible (Navier)-Stokes codes. The 3D lid-driven cavity test case deals with a flow in a unit box with five solid boundaries and moving lid as the top boundary, moving with constant velocity equal to minus one in $x$-direction. Especially the two line singularities make the lid-driven cavity problem a challenging test case.

\begin{figure}[htbp]
\centering
\includegraphics[width=1.\textwidth]{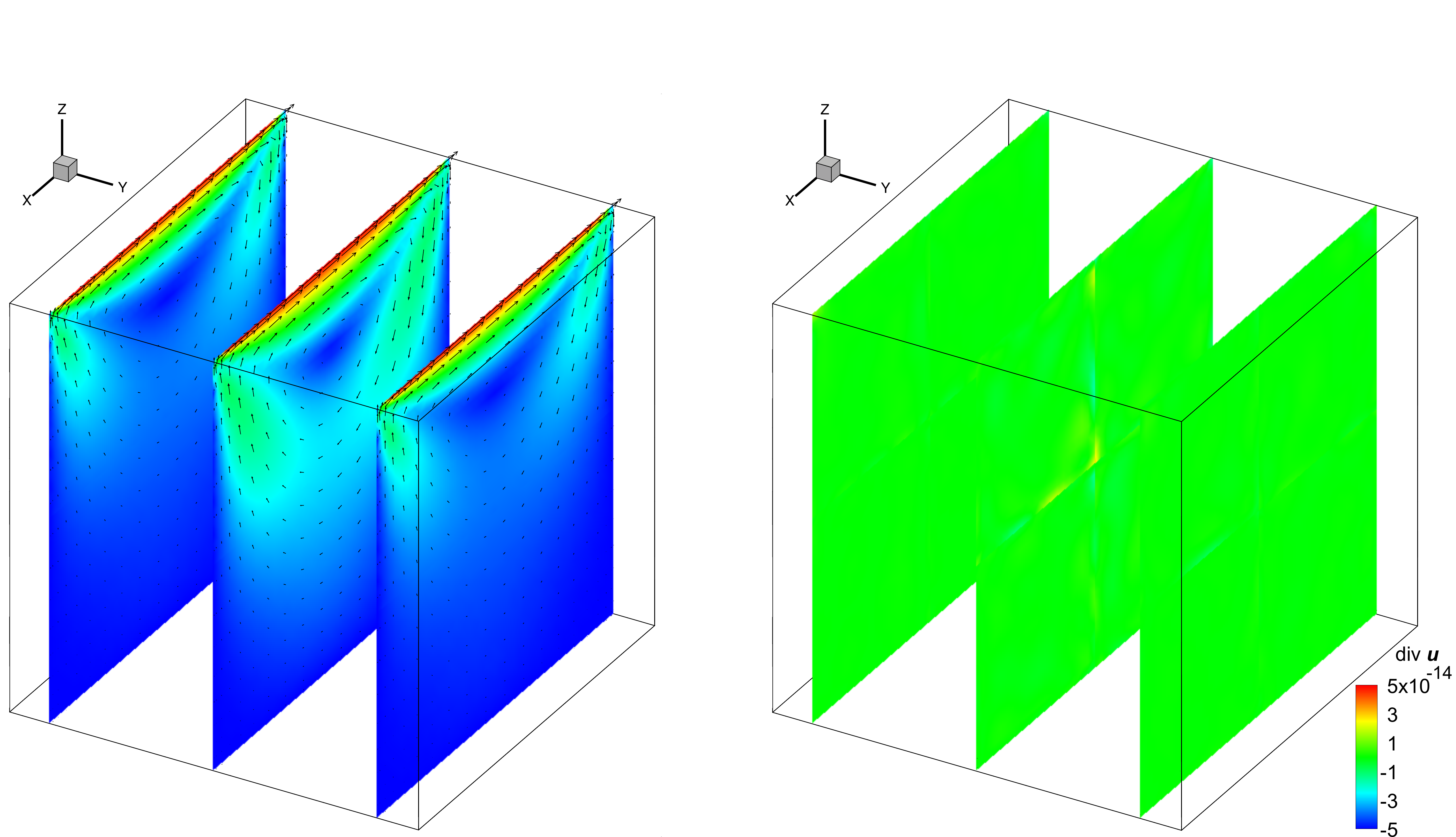}
\caption{Left: slices of magnitude of the velocity field of a three dimensional lid-driven cavity Stokes problem obtained on a $2\times2\times2$ element mesh with $N=8$. Right: slices of the  divergence of velocity. Is confirms a divergence-free velocity field.}
\label{fig:3Dldcdivfree3D}
\end{figure}

The left plot in \figref{fig:3Dldcdivfree3D} shows slices of the magnitude of the velocity field in a three dimensional lid-driven cavity Stokes problem, obtained on a $2\times2\times2$ element mesh, where each element contains a Gauss-Lobatto mesh of $N=8$. The slices are taken at 10\%, 50\% and 90\% of the y-axis. The right plot in \figref{fig:3Dldcdivfree3D} shows slices of divergence of the velocity field. \figref{fig:3Dldcdivfree3D} confirms that the mixed mimetic spectral element method leads to an accurate result with a divergence-free solution.

The second testcase shows the optimal convergence behavior for a 2D Stokes problem with no-slip boundary conditions. The testcase originates from a recent paper by Arnold et al \cite{arnold2011}, where sub-optimal convergence is shown and proven for no-slip boundary conditions when using Raviart-Thomas elements. Since Raviart-Thomas elements are the most popular $H(\mathrm{div,\Omega})$ conforming elements, we compare our method to these results.

\begin{figure}[htbp]
\centering
\includegraphics[width=1.\textwidth]{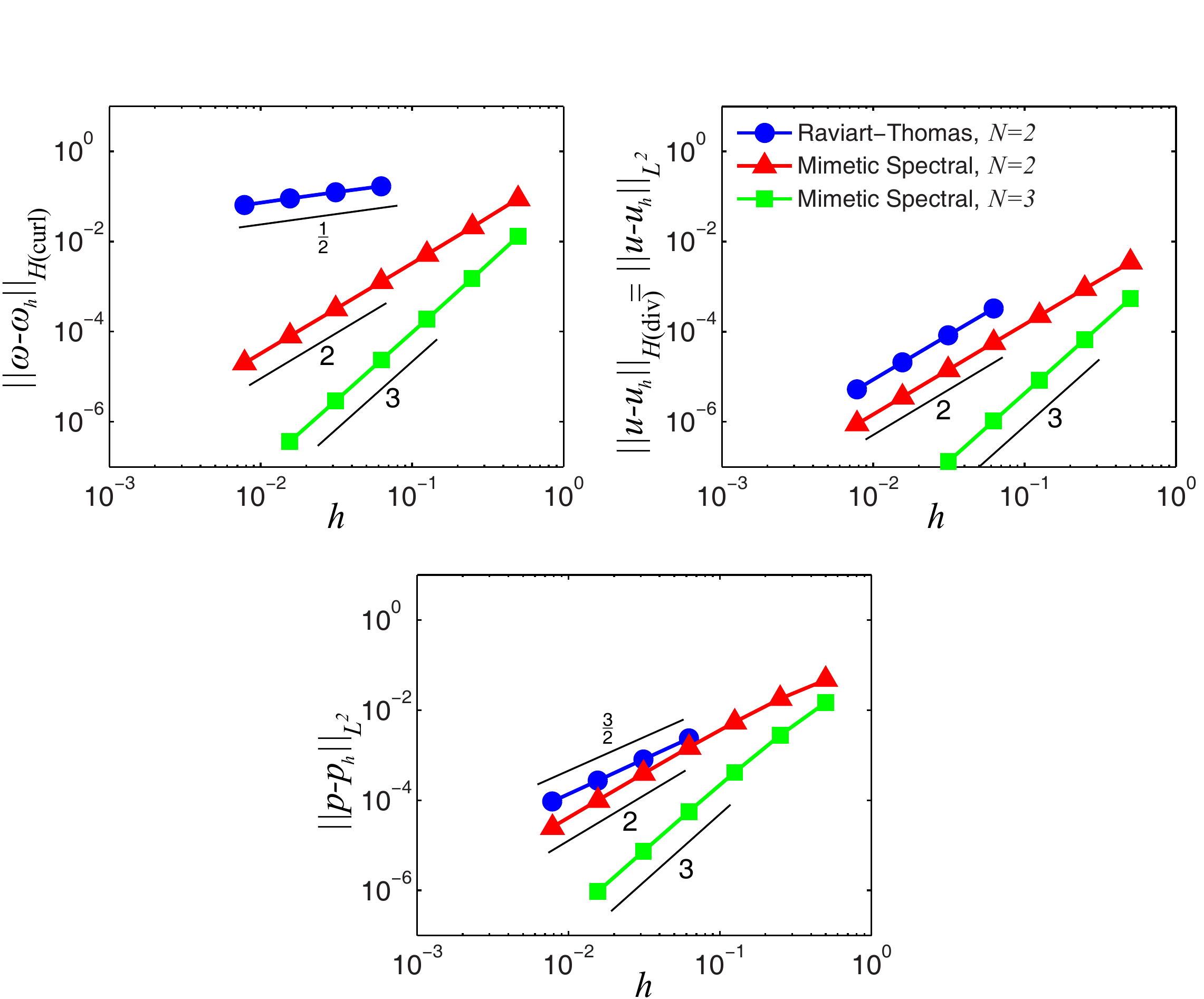}
\caption{Comparison of the $h$-convergence between Raviart-Thomas and Mimetic spectral element projections for the 2D Stokes problem with no-slip boundary conditions.}
\label{fig:afgcase3}
\end{figure}

\figref{fig:afgcase3} shows the results of the Stokes problem on a unit square with velocity and pressure fields given by $\vec{u}=\big[-2x^2(x-1)^2y(2y-1)(y-1),2y^2(y-1)^2x(2x-1)(x-1)\big]^T$, $p=(x-\tfrac{1}{2})^5+(y-\tfrac{1}{2})^5$.
While for velocity both methods show optimal convergence, for pressure a difference of $\tfrac{1}{2}$ is noticed in the rate of convergence and for vorticity a difference in rate of convergence of $\tfrac{3}{2}$ is revealed.


\bibliographystyle{abbrv}
\bibliography{./literature}
\end{document}